\documentclass[letterpaper,english,prb]{revtex4}
\usepackage[T1]{fontenc}
\usepackage[latin1]{inputenc}
\usepackage{amsmath}
\usepackage{graphicx}
\usepackage{amssymb}

\makeatletter

\providecommand{\tabularnewline}{\\}
\newcommand{\lyxdot}{.}

\usepackage{babel}
\makeatother
\begin{document}

\title{Structure and Aggregation of a Helix-Forming Polymer }

\author{James E. Magee}

\author{Zhankai Song}

\author{Robin A. Curtis}

\author{Leo Lue}

\affiliation{School of Chemical Engineering and Analytical Science, The University
of Manchester, PO Box 88, Sackville Street, Manchester, M60 1QD, United
Kingdom}

\begin{abstract}
We have studied the competition between helix formation and aggregation
for a simple polymer model. We present simulation results for a system
of two such polymers, examining the potential of mean force, the balance
between inter and intramolecular interactions, and the promotion or
disruption of secondary structure brought on by the proximity of the
two molecules. In particular, we demonstrate that proximity between
two such molecules can stabilize secondary structure. However, for
this model, observed secondary structure is not stable enough to prevent
collapse of the system into an unstructured globule. 
\end{abstract}
\maketitle

\section{\label{sec:Introduction} Introduction}

Aggregation of partially folded proteins is a problem in many bio-related
fields. In bioprocessing operations, mechanical and/or environmental
stresses lead to protein unfolding which may result in irreversible
aggregation and loss of protein activity.\cite{Proposal1} Another
problem in bioprocessing is the recovery of partially-folded recombinant
proteins, produced in densely packed {}``inclusion bodies''. Recovery
of these proteins requires dissolving the bodies in denaturant and
then slowly diluting the solution, minimizing unwanted aggregation
such that the protein folds to the native state. Finally, protein
aggregation has been implicated in the origin of amyloid fibrils,
which have been linked to many different diseases.\cite{proposal12,proposal13,proposal14}
Understanding the process of aggregation, and determining the conditions
under which it will take place, is therefore a matter of some importance.

The competition between protein folding and protein aggregation is
believed to be determined by the balance of inter and intramolecular
forces. Protein folds are stabilized by strong intramolecular interactions
between hydrophobic and hydrogen bonding groups in the protein interior.
Aggregation of folded proteins is of necessity weak, mediated by interactions
between heavily solvated charged and polar surface groups. However,
at high temperatures, or upon addition of denaturant, the fold becomes
less stable. This structural disruption causes hydrogen bonding and
hydrophobic groups to become exposed, providing {}``sticky'' sites
for aggregation. The resulting competition between structure formation
and aggregation may hold the key to understanding the process of aggregation.

The details of this process are difficult to experimentally resolve,
since partially folded intermediates have very low solubilities. As
such, computational approaches are extremely useful. Detailed, all-atom
models have been used to study the process of aggregation.\cite{Arm05,robinpaper,aggregation1,aggregation2,aggregation3}
However, such models are computationally intensive, and details of
the force field used can cloud underlying behavior. Simple {}``reduced
models'' that can produce protein-like secondary structure\cite{HallModelA,HallModelB,HallModelC,HallModelFibrillization,MaritanString,ThickPolymers,BuhotHalperin2002,Muthukumar1996,Varshneyetal2004}
are therefore useful as a parallel approach to the problem.

In a previous paper,\cite{SHFP} we proposed an isotropic polymer
model which can form helical secondary structure. This model consists
of a chain of overlapping spherical beads, and is related to the well-studied
tangent-sphere square-well polymer chain model. Low temperature transitions
to {}``crystalline'' ordered states have been observed for tangent
sphere polymers \cite{homofolding1,homofolding2,HPfolding} Further,
interactions between pairs of tangent-sphere polymers are well studied.\cite{hallPMFs}

In this paper, we extend the study of the overlapping bead model by
examining the behavior of isolated pairs of such polymers. In particular,
we examine how the formation of helical structure affects the interactions
between the two molecules and influences the structure of the resulting
complex. The remainder of this paper is organized as follows. In section
\ref{sec:Method}, we describe the simulation methodology and the
particulars of the model. In section \ref{sec:Results}, we present
the results from the simulations, and in section \ref{sec:Conclusions},
we present our conclusions.

\section{\label{sec:Method} Simulation details}

The polymer model we consider consists of a linear chain of $N$ impenetrable
spheres of diameter $\sigma$. The distance between bonded spheres
is restricted to vary between $0.9l$ and $1.1l$, where $l$ is the
nominal bond length. Directly bonded spheres may overlap (i.e., $\sigma>l$),
although non-bonded spheres may not. In addition to excluded volume
interactions, the spheres also interact via a square-well potential.
The interaction potential $u(r)$ between the spheres is given by:
\begin{equation}
u(r)=\left\{ \begin{array}{ccc}
\infty & \mathrm{for} & r<\sigma\\
-\epsilon & \mathrm{for} & \sigma<r<\lambda\sigma\\
0 & \mathrm{for} & \lambda\sigma<r\end{array}\right.\label{eq:sqwell}\end{equation}
where $r$ is the distance between the centers of the spheres, $\epsilon$
is the strength of the square-well interaction, and $\lambda$ characterizes
the width of the square-well interaction. In this work, $\lambda=1.5$.
Since this potential is pairwise, the interaction energy of a system
of polymers can be decomposed into an intramolecular energy $E_{intra}$
(interactions between monomers on the same molecule) and an intermolecular
energy $E_{inter}$ (interactions between monomers on different molecules).
At $\sigma/l=1.0$, the model reduces to the traditional tangent sphere
polymer model; interactions between pairs of such polymers has been
studied extensively elsewhere.\cite{hallPMFs,Striolo1,Striolo2,SWpairs1,SWpairs2}
For larger values of $\sigma/l$, we have observed helix formation,\cite{SHFP}
despite the isotropic, achiral nature of the model. While the precise
mechanism of helix formation for this model remains unclear, it has
been suggested that helices are in fact the optimal packing structure
for any string-like object,\cite{Maritanetal2000} in the same fashion
that optimal packing of spheres yields hexagonal or cubic close-packed
periodic lattices.

Monte Carlo simulations were performed for a pair of $N=20$ (20mer)
square-well molecules with $\sigma/l=1.3$, $1.6$, and $1.9$. Pivot,
crank-shaft, and sphere displacement moves were used to sample the
individual polymer conformations. Overall translation and rotation
moves of the polymers were performed to sample the relative positions
and orientations of the polymers. Umbrella sampling of the polymer
separation was performed with a potential of the form \begin{equation}
U(r)=a(r-b)^{2}\label{eq:umbrella}\end{equation}
where $r$ is the center of mass separation between the polymers,
and the parameter $a=40k_{B}T$ was used to sample intervals in polymer
center of mass separation. The parameter $b$ was varied from a value
of $0$ to 25$l$. The values of $b$ were initially distributed evenly
in units of $l$, and extra values were added wherever insufficient
overlap in $r$ histograms was observed. For each umbrella potential,
parallel tempering in temperature was used to enhance equilibration
of the low temperature systems. The temperatures used in the simulations
are given in Table~\ref{tab:temperatures}%
\begin{table}

\caption{\label{tab:temperatures} Temperatures used in simulations. }

\begin{centering}\begin{tabular}{|c|c|c||c|c|c|}
\hline 
Box&
$\beta\epsilon$&
$k_{B}T/\epsilon$&
Box&
$\beta\epsilon$&
$k_{B}T/\epsilon$\tabularnewline
\hline 
1&
0.18014&
5.55112&
11&
1.53610&
0.65100\tabularnewline
\hline 
2&
0.22518&
4.44089&
12&
1.76056&
0.56800\tabularnewline
\hline 
3&
0.28147&
3.55271&
13&
2.01207&
0.49700\tabularnewline
\hline 
4&
0.35184&
2.84217&
14&
2.29885&
0.43500\tabularnewline
\hline 
5&
0.43980&
2.27374&
15&
2.62467&
0.38100\tabularnewline
\hline 
6&
0.54976&
1.81899&
16&
3.00300&
0.33300\tabularnewline
\hline 
7&
0.68719&
1.45519&
17&
3.43643&
0.29100\tabularnewline
\hline 
8&
0.85899&
1.16415&
18&
3.92157&
0.25500\tabularnewline
\hline 
9&
1.07374&
0.93132&
19&
4.50450&
0.22200\tabularnewline
\hline 
10&
1.34218&
0.74506&
20&
5.12821&
0.19500\tabularnewline
\hline
\end{tabular}\par\end{centering}
\end{table}
. For each umbrella potential, the multiple histogram method was used
to combine simulation data at different temperatures.

Each production simulation run consisted of a total of $10^{6}$ parallel
tempering (box swap) moves. Between each of these attempted moves,
$80$ pivot, crank-shaft, and sphere displacement moves were attempted,
as well as $80$ molecular translation and rotation moves. Samples
of the system configuration were taken after every $10^{3}$ attempted
parallel tempering moves. For each set of conditions, an equilibration
run of $10^{5}$ attempted parallel tempering moves was performed,
followed by 10 production runs. The reported results are averages
over the production runs, and the uncertainties were estimated by
the jackknife method.

We are particularly interested in the influence of intermolecular
interactions on the helical structure which has been observed for
this model in isolation.\cite{SHFP} As such, structures are characterized
through the torsion $\tau$. For a continuous space curve $\mathbf{R}(s)$,
the torsion is defined as:

\begin{equation}
\tau(s)=\frac{\left(\mathbf{R}'(s)\times\mathbf{R}''(s)\right)\cdot\mathbf{R}'''(s)}{\left|\mathbf{R}'(s)\times\mathbf{R}''(s)\right|^{2}}\label{eq:tau}\end{equation}

\noindent where primes denote derivatives with respect to $s$. For
the polymer chain, the required derivatives are calculated using a
central difference approximation: \begin{align}
\mathbf{R}_{i}' & =\frac{1}{2}(\mathbf{R}_{i+1}-\mathbf{R}_{i-1})\nonumber \\
\mathbf{R}_{i}'' & =\mathbf{R}_{i+1}-2\mathbf{R}_{i}+\mathbf{R}_{i-1}\nonumber \\
\mathbf{R}_{i}''' & =\frac{1}{2}\left(\mathbf{R}_{i+2}-2\left(\mathbf{R}_{i+1}-\mathbf{R}_{i-1}\right)-\mathbf{R}_{i-2}\right)\label{eq:derivs}\end{align}
 where $\mathbf{R}_{i}$ is the vector position of monomer $i$. An
instantaneous molecular conformation is characterized by the configurational
average torsion $\bar{\tau}$, given by: \begin{equation}
\bar{\tau}=\sum_{i=3}^{N-2}\tau_{i}/\left(N-4\right)\label{eq:taubar}\end{equation}

\noindent where $\tau_{i}$ is the torsion at monomer $i$, defined
on $i\in\left[3,N-2\right]$, and the bar denotes an instantaneous
configurational average. Chiral (torsional) symmetry breaking is used
to detect helix structure formation; helix formation is associated
with a probability distribution $p\left(\bar{\tau}\right)$ which
is bimodal, symmetric, and with $p\left(\bar{\tau}=0\right)\approx0$.
Combined with observation of helical snapshot configurations, this
provides strong evidence for helicity. Structures are also characterized
by the absolute of the mode of $p\left(\bar{\tau}\right)$, denoted
$\bar{\tau}_{max}$.

It is also instructive to consider statistical correlations between
properties of the molecules. The correlation between observables $\mathcal{A}$
and $\mathcal{B}$ is given by:

\begin{equation}
\mathrm{cor}\left(\mathcal{A},\mathcal{B}\right)=\frac{\left(\left\langle \mathcal{AB}\right\rangle -\left\langle \mathcal{A}\right\rangle \left\langle \mathcal{B}\right\rangle \right)^{2}}{\left(\left\langle \mathcal{A}^{2}\right\rangle -\left\langle \mathcal{A}\right\rangle ^{2}\right)\left(\left\langle \mathcal{B}^{2}\right\rangle -\left\langle \mathcal{B}\right\rangle ^{2}\right)}\label{eq:cor}\end{equation}

\noindent where angle brackets denote ensemble averaging. A correlation
of zero indicates that the two variables are completely uncorrelated,
while a correlation of one indicates perfect linear correlation. We
measure correlations between the intramolecular energies of the individual
molecules $\mathrm{cor}\left(E_{intra}^{(1)},E_{intra}^{(2)}\right)$,
the torsions of the individual molecules $\mathrm{cor}\left(\bar{\tau_{1}},\bar{\tau}_{2}\right)$,
and the orientations of the end-to-end bond vectors, $\left\langle \left|\mathbf{\hat{r}}_{e(1)}\cdot\hat{\mathbf{r}}_{e(2)}\right|\right\rangle $.
The last quantity should take a value of one if the end-to-end vectors
of the molecules are aligned, zero if the end-to-end vectors are perpendicular,
and a half if the orientations are uncorrelated. The absolute value
of the dot product is taken because the molecules have no directionality;
antiparallel orientations are equivalent to parallel orientations.
For unstructured globules, we expect this quantity to be one half
(since the end-to-end vector should be randomly oriented), whereas
for helical structures, the end-to-end vectors are a reasonable approximation
for the helical axes, and the dot product provides a measure of the
orientational correlation of the helices.

The pair potential of mean force (PMF) $\psi(r)$ is estimated from
the observed histograms $h(r)$ of separation between molecules. For
a given simulation, this is given by: \begin{equation}
\psi(r)=-k_{B}T\ln\left(h(r)/r^{2}\right)-U(r)+C\label{eq:PMF}\end{equation}
where $k_{B}$ is the Boltzmann constant and $C$ is an unknown energy
offset, arising from the incomplete normalization of probability histograms
from simulation. The $U(r)$ term removes the bias introduced by the
umbrella potentials. Since solvent is implicit, $\psi(r)=0$ at separations
for which the polymers can have no direct interaction (i.e.. $r/l>\left(N-1\right)+\lambda\sigma/l$).
For simulations sampling these large values of $r$, it is possible
to fix $C$ by this criterion. Results for neighboring umbrellas are
then {}``patched'' together by finding values for $C$ which minimize
the total statistical uncertainty in the combined results. 

The last quantity of interest is the second virial coefficient $B_{2}$,
which may be calculated from the PMF via:

\begin{equation}
B_{2}=2\pi\int\left(1-\exp\left(-\beta\psi(r)\right)\right)r^{2}\mathrm{d}r\label{eq:B2}\end{equation}

\noindent Positive values for $B_{2}$ indicate an overall repulsive
force acting between the molecules; negative values indicate an overall
attractive force.

\section{\label{sec:Results}Results and discussion}

\begin{figure}
\includegraphics{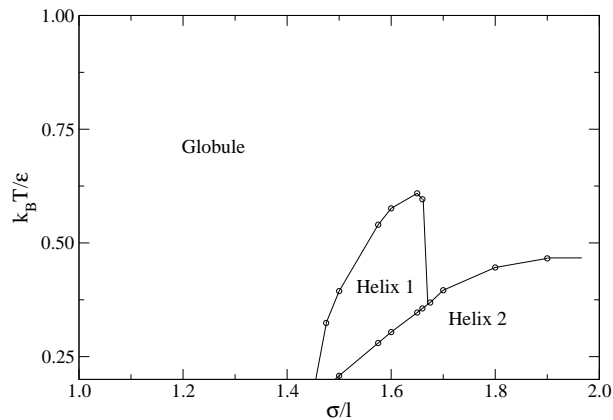}

\caption{\label{fig:singlephdat} Single molecule phase diagram.}
\end{figure}

In previous work,\cite{SHFP} we have studied this polymer model for
isolated 20mers, down to a temperature of $k_{B}T/\epsilon=0.2$.
It was found that, at sufficiently low temperatures, these polymers
may form helical structures, depending upon the value for $\sigma/l$.
Further, two separate helical phases were observed over the range
$1.475\leq\sigma/l\leq1.66$. For reference, the phase diagram of
the single polymer system is reproduced in Fig.~\ref{fig:singlephdat}.%
 In this work, we study the behavior of an isolated pair of polymer
molecules with $\sigma/l=1.3$, $1.6$ and $1.9$.

\subsection{$\sigma/l=1.6$}

We begin with a description of the results for the $\sigma/l=1.6$
system, which best demonstrates behaviors which also occur in the
$\sigma/l=1.3$ and $\sigma/l=1.9$ systems. Results are presented
across a range of separations at four representative temperatures;
$k_{B}T/\epsilon=0.255$, for which isolated polymers are in the helix
2 phase, $k_{B}T/\epsilon=0.435$, for which isolated polymers are
in the helix 1 phase, $k_{B}T/\epsilon=0.651$, just above the helix
1-globule transition temperature, and $k_{B}T/\epsilon=1.164$, at
which the polymer is well into the globule phase.

\begin{figure}
\begin{centering}\includegraphics[clip]{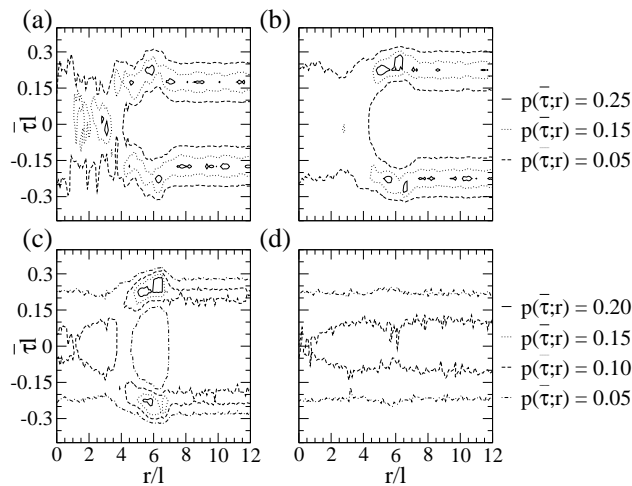}\par\end{centering}

\caption{\label{fig:poftau1.6} Contour plots of the estimated probability
distributions $p\left(\bar{\tau};r\right)$ for $\sigma/l=1.6$ at
(a) $k_{B}T/\epsilon=0.255$, (b) $k_{B}T/\epsilon=0.435$, (c) $k_{B}T/\epsilon=0.651$,
and (d) $k_{B}T/\epsilon=1.164$. }
\end{figure}

\begin{figure}
\includegraphics[clip]{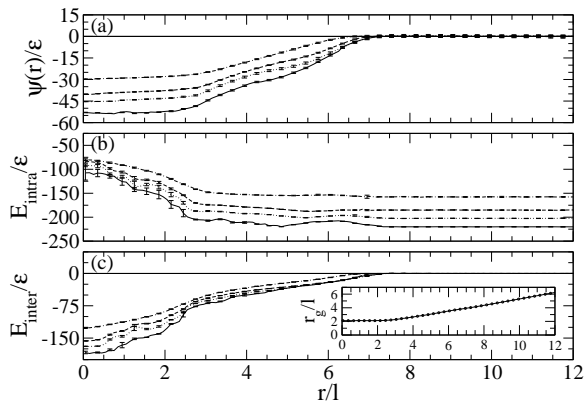}

\caption{\label{fig:potn1.6} Variation of thermodynamic properties with molecular
separation for $\sigma/l=1.6$: (a) potential of mean force, (b) intramolecular
interaction energy, and (c) intermolecular interaction energy. Solid
lines denote $k_{B}T/\epsilon=0.255$, dotted lines denote $k_{B}T/\epsilon=0.435$,
dashed lines denote $k_{B}T/\epsilon=0.651$, and dot-dashed lines
denote $k_{B}T/\epsilon=1.164$. Inset: Root mean square radius of
gyration $r_{g}/l$ for $k_{B}T/\epsilon=0.255$. Errors are less
than symbol size. Equivalent data at higher temperatures are indistinguishable
at this scale.}
\end{figure}

\begin{figure}
\includegraphics[clip]{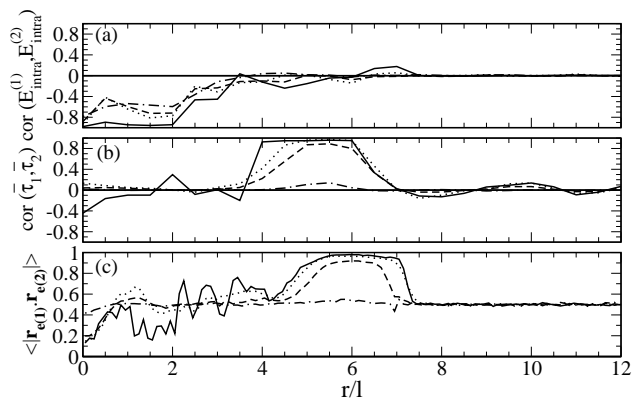}

\caption{\label{fig:1.6corr} Intermolecular correlations for $\sigma/l=1.6$
between: (a) intramolecular energies, (b) conformational torsions,
and (c) end-to-end bond vectors. Solid lines denote $k_{B}T/\epsilon=0.255$,
dotted lines denote $k_{B}T/\epsilon=0.435$, dashed lines denote
$k_{B}T/\epsilon=0.651$, and dot-dashed lines denote $k_{B}T/\epsilon=1.164$.}
\end{figure}

In order to illustrate the structural behavior of the interacting
molecules, contour plots of $p\left(\bar{\tau};r\right)$ against
$r$ are displayed in Fig.~\ref{fig:poftau1.6}.%
 The thermodynamic behavior is shown in Fig.~\ref{fig:potn1.6}.%
 The potential of mean force is given in Fig.~\ref{fig:potn1.6}(a),
the mean intramolecular interaction energy $\left\langle E_{intra}\right\rangle $
is given in Fig.~\ref{fig:potn1.6}(b), and the variation of the
intermolecular interaction energy $\left\langle E_{inter}\right\rangle $
is given in Fig.~\ref{fig:potn1.6}(c). The correlations between
the intramolecular energies $\mathrm{cor}\left(E_{1},E_{2}\right)$,
the mean torsions $\mathrm{cor}\left(\bar{\tau}_{1},\bar{\tau}_{2}\right)$,
and the orientation of the end-to-end distance vectors $\left\langle \mathbf{r}_{e(1)}\cdot\mathbf{r}_{e(2)}\right\rangle $,
for the molecules are shown in Fig.~\ref{fig:1.6corr}.%

The lowest representative temperature ($k_{B}T/\epsilon=0.255$) is
less than the helix 1-helix 2 transition temperature, and the results
are typical of this region. In Fig.~\ref{fig:poftau1.6}(a), it can
be seen that at separations $r/l\gtrsim7.5$, $p(\bar{\tau};r)$ is
independent of $r$, bimodal and approximately zero at $\bar{\tau}=0$.
In Fig.~\ref{fig:potn1.6} (solid line), it can be seen that, at
these large separations, $\psi(r)$ and $\left\langle E_{inter}\right\rangle $
are approximately zero, and $\left\langle E_{intra}\right\rangle $
appears $r$-independent. Finally, as shown in Fig.~\ref{fig:1.6corr}
(solid line), all three measured correlators are approximately zero
for large separations. This behavior indicates that, for $r/l\gtrsim7.5$,
the two molecules do not significantly interact with each other, existing
independently in the helix 2 state. There is a weak interaction, since
observed $\left\langle E_{inter}\right\rangle $ is non-zero, with
magnitude greater than observed error, for $r/l\lesssim8.5$; however,
these interactions do not appear strong enough to significantly perturb
the isolated equilibrium state.

The interactions between the molecules begin to perturb their individual
structures at $r/l\approx7.5$. In the range $4\lesssim r/l\lesssim7.5$,
$p\left(\bar{\tau};r\right)$ remains bimodal, however, the positions
of the peaks, $\pm\bar{\tau}_{max}$ depend upon separation. As shown
in Fig.~\ref{fig:potn1.6}, over the same range, $\psi(r)$ and $\left\langle E_{inter}\right\rangle $
are monotonically increasing functions of $r$, while $\left\langle E_{intra}\right\rangle $
remains approximately constant. Further, the torsions and end-to-end
vectors of the molecules become strongly correlated in this region,
while intramolecular energies remain uncorrelated (see Fig.~\ref{fig:1.6corr}).

\begin{figure}
\begin{centering}\includegraphics{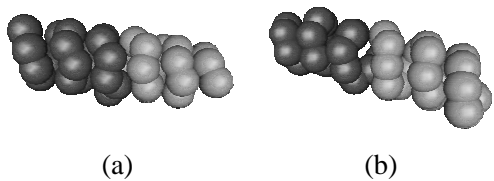}\par\end{centering}

\caption{\label{fig:gallery} Snapshot configuration for (a) $\sigma/l=1.6$,
$k_{B}T/\epsilon=0.255$, umbrella potential centered at $b/l=5$,
and (b) $\sigma/l=1.6$, $k_{B}T/\epsilon=0.651$, umbrella potential
centered at $b/l=5.6$.}
\end{figure}

The bimodality of the torsion distributions, as well as the approximately
constant value for the intramolecular energies, indicates that the
molecules maintain helical configurations within this range. A snapshot
configuration of two helical molecules taken at this temperature is
presented in Fig.~\ref{fig:gallery}(a).%
 The observed correlations show that the helices have a strong preference
to take on the same handedness, and lie end-to-end along the same
axis.

The average end-to-end distance for the unperturbed polymer at this
temperature\cite{SHFP} is $r_{e}/l\approx5$; two completely rigid
{}``average'' helices would only begin to interact at $r/l=5.5$.
For two helices to interact at $r/l=7.5$, they must have $r_{e}/l>7$,
which corresponds to an extension of at least $27\%$. The interactions
between the molecules appear to stretch the molecules, relative to
their isolated equilibria. 

The stretching effect weakens upon closer approach between the molecules;
at separations $4\lesssim r/l\lesssim5$, $\bar{\tau}_{max}$ approximately
returns to its value at large separation. Further, $\left\langle |\mathbf{r}_{e(1)}\cdot\mathbf{r}_{e(2)}|\right\rangle $
decreases to approximately $0.5$ at separations in this range, indicating
that the two molecules are less likely to align. For an isolated molecule
at this temperature, $r_{e}/l\approx5$; for two aligned {}``average''
helices, there would be a longitudinal compression at separations
closer than this. As such, the stretching force is eliminated, and
in order to remain helical, the helical axes must lose some orientational
correlation upon further approach. Note that the torsion correlator
remains high across this region. In the range $4\lesssim r/l\lesssim7.5$,
the intramolecular energies of the two molecules are uncorrelated. 

Once the two molecules approach closer than $r/l\approx4$, $p\left(\bar{\tau};r\right)$
becomes unimodal (see Fig.~\ref{fig:poftau1.6}); that is, the structures
become achiral, and helicity is lost. This behavior is marked by a
small discontinuity in total energy (data not shown) at $r/l=4.15$
and a sudden loss of torsional correlations between the molecules.
For $2.5\lesssim r/l\lesssim4$, $\psi(r)$ and $\left\langle E_{inter}\right\rangle $
increase monotonically with $r$, $\left\langle E_{intra}\right\rangle $
remains approximately constant, and no significant correlation exists
between the intramolecular energies. The intermolecular interactions
in this range are strong enough to fully disrupt the helical structures,
causing both molecules to adopt individual unstructured globule configurations.
We suggest that the lack of intramolecular energy correlations and
the approximately constant intramolecular energy indicate that, at
these separations, the complex consists of two globules with only
limited interpenetration. The steady decrease in $\left\langle E_{inter}\right\rangle $
with decreasing $r$ suggests that the two globules deform upon approach
to give a larger contacting interface.

At $r/l=2.5$, however, there is a large discontinuity in $\left\langle E_{inter}\right\rangle $
and in $\left\langle E_{intra}\right\rangle $. Further, the slope
of $\left\langle E_{intra}\right\rangle $ versus separation changes
at this separation; for $r/l>2.5$, the average intramolecular energy
is approximately constant (with respect to $r$), however, for $r/l<2.5$,
it is a decreasing function of $r$. For $r<2.5$, there is a negative
correlation between the intramolecular energies. This shift has no
obvious signature in $p\left(\bar{\tau};r\right)$. The inset to Fig.~
\ref{fig:potn1.6} shows that, below $r/l\approx2.5$, the radius
of gyration of the system becomes approximately constant ($r_{g}/l\approx2.1$);
further approach does not lead to contraction of the complex. The
\emph{total} energy of the system (i.e., $E=E_{intra}+E_{inter}$)
also becomes approximately constant with respect to $r$ at these
short separations (data not shown). This suggests increasing interpenetration
between the two molecules, with intramolecular contacts being replaced
on a one-to-one basis by intermolecular contacts. Since $\psi(r)$
is also approximately constant for $r/l<2.5$, this implies that the
entropy of the system also remains approximately constant at these
separations. Strong anticorrelation of intramolecular energies, indicated
by large negative values for $\mathrm{corr}\left(E_{1},E_{2}\right)$,
shows that any {}``swelling'' (loss of intramolecular contacts)
of one molecule must be matched by an equivalent {}``collapse''
(gain in intramolecular contacts) of the other. These data are consistent
with two interpenetrated molecules in a collapsed, unstructured globule
state.

It is unclear why, for separations $2.5<r/l\lesssim4$, the two molecules
act like immiscible droplets, whereas for $r/l<2.5$, the molecules
can interpenetrate; nor is it clear why the transition between the
two is so sharp.

Qualitatively similar behavior is seen when $k_{B}T/\epsilon=0.435$,
which is in (and typical of) the helix 1 phase for an isolated polymer.
Once again, at large separations, molecules adopt two independent,
uncorrelated helical structures. Upon closer approach, interactions
once again cause correlation between the helical axis and torsions
and cause stretching forces which lead to extended helical configurations.
Closer still, correlations and helical structure are suddenly lost,
and the system appears to form two contacting but distinct unstructured
globules. Finally, interactions cause the two molecules to interpenetrate,
forming a single globule, which again is marked by a broad, flat minimum
in $\psi(r)$.

Increasing the temperature to $k_{B}T/\epsilon=0.651$, there is unexpected
behavior. This temperature is above that of the helix 1-globule transition,
and, consequently, at large separations the molecules adopt unstructured
globule configurations, with no chiral symmetry breaking. However,
at $r/l=6.8$, there is a small discontinuity in the total energy
(not shown), and $p\left(\bar{\tau};r\right)$ changes from a broad,
unimodal distribution to a bimodal distribution. That is, upon approach,
interactions between the two molecules cause chiral symmetry breaking
and helix formation, at a temperature above that where helical configurations
would normally be stable. A snapshot configuration is shown in Fig.~\ref{fig:gallery}(b).
The character of this pair of helices is very similar to the helices
seen for the same separations at lower temperatures (see Fig.~\ref{fig:gallery}(a)),
with similar torsion values and correlation in helical axes and torsions.
We suggest that attractive intermolecular interactions between the
globules produce a stretching force, which has previously been shown
to stabilize helix formation.\cite{stretching} As the two molecules
continue to approach each other, there is once again a loss of helical
structure and intermolecular correlations to two unmixed globules,
followed by interpenetration and mixing.

Finally, consider the temperature $k_{B}T/\epsilon=1.164$, which
is above that of the helix-globule transition for isolated polymers.
The torsion distribution $p\left(\bar{\tau};r\right)$ exhibits a
broad, unimodal structure for all separations; this system shows no
chiral symmetry breaking, and no helix formation. There are no observed
correlations between the torsions or the end-to-end vectors of the
molecules. There is an apparent transition between separate and interpenetrated
globules, marked by a slight discontinuity in total energy, changes
in the slope of $\psi(r)$ and the inter- and intramolecular energies,
and sudden onset of strong anticorrelations between intra-molecular
energies. It is also interesting to note that the potential of mean
force at this temperature is qualitatively similar to those for the
lower temperatures; it appears that the presence or absence of helical
structure has little effect on the character of the PMF.

\begin{figure}
\includegraphics[clip]{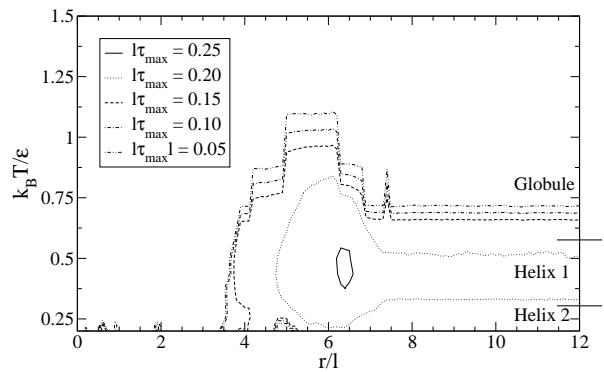}

\caption{\label{fig:1.6tau} Contour plot of the absolute mode $\tau_{max}$
of $p\left(\bar{\tau};r,T\right)$ for $\sigma/l=1.6$. The positions
of the globule - helix 1 and helix 1 - helix 2 transitions are marked
on the right hand side. Note that non-zero $\tau_{max}$ is not sufficient
for chiral symmetry breaking; see discussion in Section \ref{sec:Method}
and Ref. 19.}
\end{figure}

A contour plot of $\bar{\tau}_{max}$ as a function of $k_{B}T/\epsilon$
and $r$ is shown in Fig.~\ref{fig:1.6tau}.%
 Note that regions of non-zero $\bar{\tau}_{max}$ do not necessarily
indicate helical configurations; as pointed out in our previous work
\cite{SHFP} for single molecules, there exists a region just above
the helix transition where $p(\bar{\tau})$ is double peaked but has
a non-zero value around $\bar{\tau}=0$, and as such is not truly
symmetry breaking. However, this contour plot does schematically show
the increase in stability of helical structures at intermediate separations,
as well as the loss of helical structure at close separation.

\subsection{$\sigma/l=1.9$}

\begin{figure}
\includegraphics[clip]{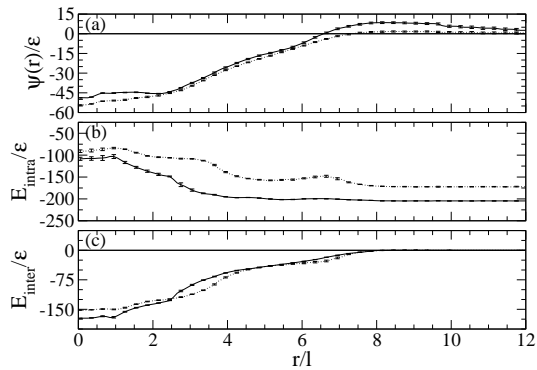}

\caption{\label{fig:potn1.9} Variation of thermodynamic properties with molecular
separation for $\sigma/l=1.9$: (a) potential of mean force, (b) intramolecular
interaction energy, and (c) intermolecular interaction energy. Solid
lines denote $k_{B}T/\epsilon=0.651$ and dotted lines denote $k_{B}T/\epsilon=1.455$. }
\end{figure}

\begin{figure}
\includegraphics[clip]{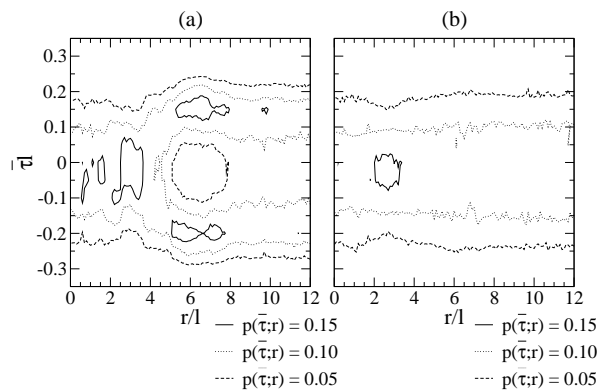}

\caption{\label{fig:poftau1.9} Contour plots of the estimated probability
distributions $p\left(\bar{\tau};r\right)$ for $\sigma/l=1.9$ at
(a) $k_{B}T/\epsilon=0.651$, and (b) $k_{B}T/\epsilon=1.455$.}
\end{figure}

\begin{figure}
\includegraphics[clip]{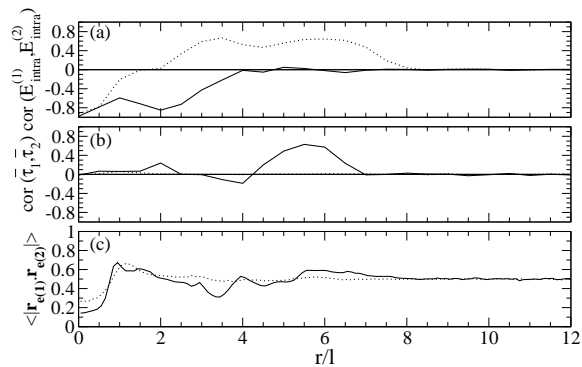}

\caption{\label{fig:1.9corr} Intermolecular correlations for $\sigma/l=1.9$
between: (a) intramolecular energies, (b) conformational torsions,
and (c) end-to-end bond vectors. Solid lines denote $k_{B}T/\epsilon=0.651$,
dotted lines denote $k_{B}T/\epsilon=1.455$. }
\end{figure}

\begin{figure}
\includegraphics[clip]{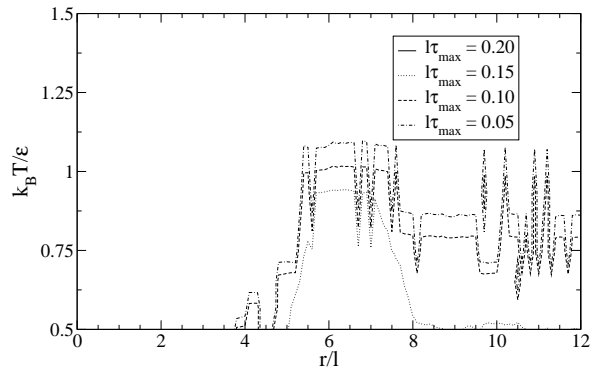}

\caption{\label{fig:1.9tau} Contour plot of the absolute mode $\tau_{max}$
of $p\left(\bar{\tau};r\right)$ for $\sigma/l=1.9$. All temperatures
shown are above the helix 2 - globule transition temperature. }
\end{figure}

Potential, torsion and correlation data for the $\sigma/l=1.9$ system
are shown in Figs.~\ref{fig:potn1.9}, \ref{fig:poftau1.9} and \ref{fig:1.9corr}.%
 A plot of $\bar{\tau}_{max}$ against separation and temperature
is shown in Fig.~\ref{fig:1.9tau}. Once again, we consider representative
temperatures: this time, $k_{B}T/\epsilon=0.651$, typical of the
region just above the helix 2 - globule and $k_{B}T/\epsilon=1.455$,
typical of higher temperatures. We were not able to fully equilibrate
the two molecules at temperatures below the helix 2 - globule transition.

The behavior at the lower temperature $k_{B}T/\epsilon=0.651$ is
qualitatively similar to the situation for the $\sigma/l=1.6$ system
at $k_{B}T/\epsilon=0.651$. The torsion histograms in Fig.~\ref{fig:poftau1.9}(a)
reveal a range of separations over which helical structures are observed
at temperatures higher than found for isolated molecules. This range
is associated with a correlation between the torsions of the two molecules
(Fig.~\ref{fig:1.9corr}(b), solid line), although this correlation
is not as strong as that observed for the $\sigma/l=1.6$ system.
Interestingly, there is no significant correlation in the end-to-end
bond vectors over this range (Fig.~\ref{fig:1.9corr}(c), solid line).
At close separations, the intramolecular energies of the two molecules
become strongly anticorrelated (Fig.~\ref{fig:1.9corr}(a), solid
line). This correlation appears gradually on approach from the loss
of helical structure (at $r/l\approx4$) to a discontinuity in intramolecular
and intermolecular energies (at $r/l=2.4$). Once again, this separation
coincides with the point at which the radius of gyration (not shown)
becomes approximately constant, with $r_{g}/l\approx2.4$, and below
this separation, total energy and $\psi(r)$ (Fig.~\ref{fig:potn1.9}(a),
solid line) are also independent of $r$. Aside from the lack of orientational
correlation, the only other qualitative difference is a repulsive
{}``hump'' in $\psi(r)$ at long ranges for the $\sigma/l=1.9$
system. The maximum of this barrier $(r/l=8.25)$ is near to the separation
at which $\left\langle E_{inter}\right\rangle $ becomes non-zero
--- this barrier is, therefore, entropic in nature, corresponding
to loss of entropy of extended configurations. The long range of this
hump suggests that such extended configurations are relatively frequent
occurrences --- this is to be expected, considering the stiffness
of the $\sigma/l=1.9$ polymer. Overall, then, the results are consistent
with the interpretations offered for the $\sigma/l=1.6$ system at
the same temperature.

At $k_{B}T/\epsilon=1.455$, there is new behavior. While the torsional
behavior (Fig.~\ref{fig:poftau1.9}(b)), potential of mean force
and energies (Fig.~\ref{fig:potn1.9}, dotted line), and the torsional
and end-to-end vector correlators (Fig.~\ref{fig:1.9corr}(b) (c),
dotted line) appear similar to those seen at high temperatures for
$\sigma/l=1.6$, there are strong internal energy correlations across
the range $2\lesssim r\lesssim8$ (Fig.~\ref{fig:1.9corr}(a), dotted
line). These correlations are observed at temperatures up to $k_{B}T/\epsilon=2.273$
- above this, the system returns to the qualitative behavior seen
for $\sigma/l=1.6$, $k_{B}T/\epsilon=1.164$. Snapshot configurations
taken across this range of temperature and separation do not show
any remarkable common features; the system appears to be exploring
a large ensemble of unstructured configurations.

The positive intramolecular energy correlation indicates that swelling
of one polymer is associated with swelling of the other. We relate
this to the stretching force that we have inferred to operate between
helices over a similar range of separations. If one of the polymers
is in a particularly open (high intramolecular energy) configuration,
then it will have a larger radius. This will increase the likelihood
of interactions with the other polymer; if these interactions are
attractive, then the other polymer can only expand (at fixed intermolecular
separation). The variance in intramolecular energies for the $\sigma/l=1.9$
system in the globule phase is approximately double that observed
for the $\sigma/l=1.6$ globule phase (data not shown). This indicates
that the $\sigma/l=1.9$ globule phase has much lower internal cohesion
than the globules at smaller overlaps, which we suggest is due the
inherent stiffness of the polymer, and which allows cross interactions
to produce these cross-correlations which are not observed for the
$\sigma/l=1.6$ system.

In summary, the behavior of the $\sigma/l=1.9$ system is largely
similar to that of the $\sigma/l=1.6$ system, with destruction of
helical structure at short range, stabilization of helical structure
at intermediate range, and an apparent transition between separate
and interpenetrated globules at very close range. A summary of the
torsional behavior of the system is shown in Fig.~\ref{fig:1.9tau}.%
 Two significant differences in behavior occur. First, strong correlations
in intramolecular energies between the two molecules are observed
in the unstructured globule phase. We suggest that this does not occur
in the $\sigma/l=1.6$ system because the individual globules are
more internally cohesive, as indicated by lower intramolecular energy
fluctuations. Further, no correlations between the end-to-end bond
vectors of the helices are observed; we will return to this point
in the conclusion.

\subsection{$\sigma/l=1.3$}

\begin{figure}
\includegraphics[clip]{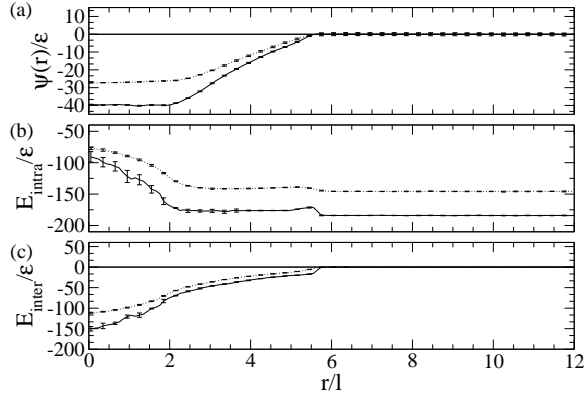}

\caption{\label{fig:potn1.3} Variation of thermodynamic properties with intermolecular
separation for $\sigma/l=1.3$: (a) potential of mean force, (b) intramolecular
interaction energy, and (c) intermolecular interaction energy. Solid
lines denote $k_{B}T/\epsilon=0.255$, and dotted lines denote $k_{B}T/\epsilon=0.651$. }
\end{figure}

\begin{figure}
\includegraphics[clip]{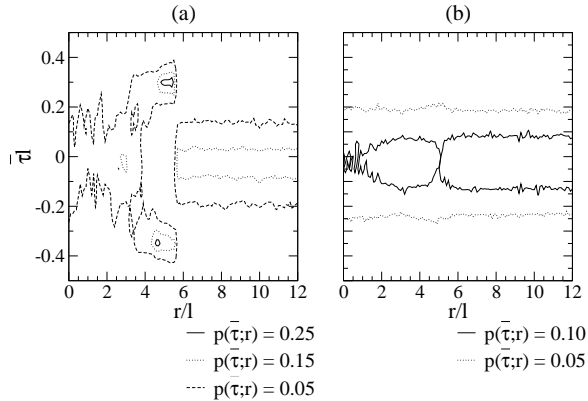}

\caption{\label{fig:poftau1.3} Contour plots of the estimated probability
distributions $p\left(\bar{\tau};r\right)$ for $\sigma/l=1.3$ at
(a) $k_{B}T/\epsilon=0.255$, and (b) $k_{B}T/\epsilon=0.651$.}
\end{figure}

\begin{figure}
\includegraphics[clip]{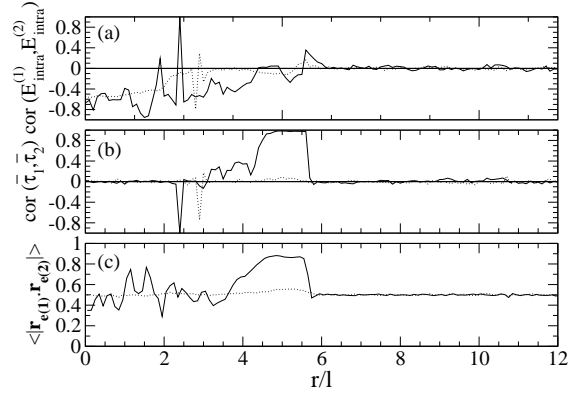}

\caption{\label{fig:1.3corr} Intermolecular correlations for $\sigma/l=1.3$
between: (a) intramolecular energies, (b) conformational torsions,
and (c) end-to-end bond vectors. Solid lines denote $k_{B}T/\epsilon=0.255$;
dotted lines denote $k_{B}T/\epsilon=0.651$.}
\end{figure}

Finally, consider square-well polymers with $\sigma/l=1.3$. Here,
we present two representative temperatures, $k_{B}T/\epsilon=0.255$
and $k_{B}T/\epsilon=0.651$. Single 20mers with $\sigma/l=1.3$ do
not display \cite{SHFP} any helical structure down to a temperature
of $k_{B}T/\epsilon=0.2$. Potential of mean force and energies are
shown in Fig.~\ref{fig:potn1.3},%
 contour plots of $p\left(\bar{\tau};r\right)$ in Fig.~\ref{fig:poftau1.3},%
 and correlations in Fig.~\ref{fig:1.3corr}.%

At the lower temperature $k_{B}T/\epsilon=0.255$, the behavior is
similar to that seen for the $\sigma/l=1.6$ and $\sigma/l=1.9$ systems
at temperatures just above the helix-globule transition. At long range,
there is no torsional symmetry breaking. This continues on approach
down to $r/l=5.65$, where there is a sudden discontinuity in intramolecular
and intermolecular energies. At this separation, $p(\bar{\tau};r)$
abruptly becomes bimodal, and the torsions and end-to-end vectors
become strongly correlated. This behavior indicates that the helix
phase is stabilized by interactions between the polymers. This stability
continues across the range $3.65\lesssim r/l\lesssim5.65$. At $r/l=3.6$,
there is a small discontinuity in the total energy, the torsion histogram
is no longer bimodal, and the torsions and orientations of the molecules
are no longer correlated. At closer separations, there is a gradual
increase in anticorrelation between the intramolecular energies up
to a discontinuity in intermolecular energy at $r/l=1.8$. Upon closer
approach, the intramolecular energies are strongly anticorrelated,
and the PMF and total system $r_{g}$ remain approximately constant.
This behavior is once again interpreted as a transition from two immiscible
polymer globules at intermediate separation to two interpenetrating
globules at close separations.

At the higher temperature $k_{B}T/\epsilon=0.651$, the behavior is
similar to that seen for the $\sigma/l=1.6$ and $\sigma/l=1.9$ systems
in the unstructured globule regime. The torsions and end-to-end vectors
remain uncorrelated at all separations, and the torsion histograms
remain unimodal. The mixing transition between the two globules appears
at close separation, marked by the onset of intramolecular energy
correlations. No strong intramolecular energy correlations are observed.
The variance in the intramolecular energies is of the order of magnitude
of that seen for the $\sigma/l=1.6$ system, which suggests that the
globules have enough internal cohesion to prevent such correlation.

\begin{figure}
\includegraphics[clip]{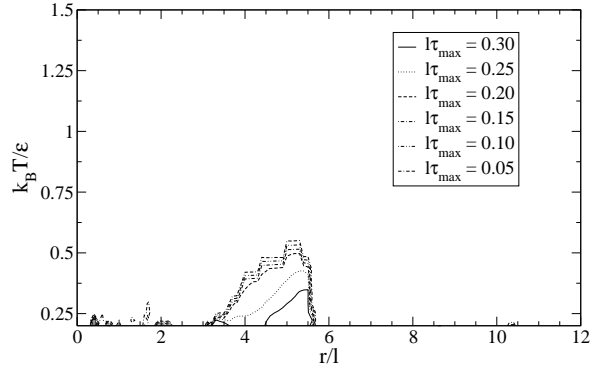}

\caption{\label{fig:1.3tau} Contour plot of the absolute mode $\tau_{max}$
of $p\left(\bar{\tau};r\right)$ for $\sigma/l=1.6$.}
\end{figure}

The torsional data are summarized in Fig.~\ref{fig:1.3tau}%
 and are similar to the data presented in Figs.~\ref{fig:1.6tau}
and \ref{fig:1.9tau} for temperatures above the helix-globule transition.

\subsection{\label{sec:Summary}Summary}

\begin{figure}
\includegraphics{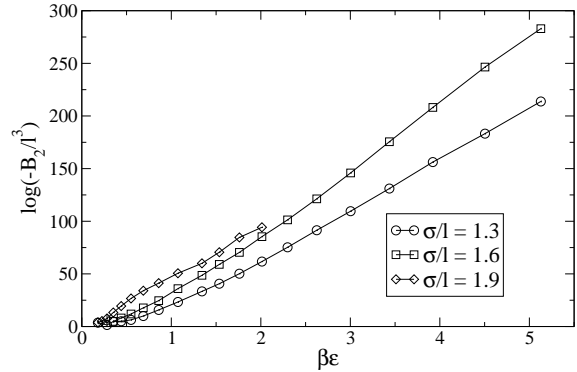}

\caption{\label{fig:B2fig} Log-reciprocal plot of minus the second virial
coefficient, $-B_{2}/l^{3}$ against temperature, $k_{B}T/\epsilon$
for the three values of $\sigma/l$ studied. Errors are smaller than
symbol size.}
\end{figure}

In summary, we have found similar behaviors for pairs of model polymers
for $\sigma/l=1.3$, $1.6$ and $1.9$. In each case, intermediate
separations between the molecules enhance helix stability, showing
paired helical configurations at temperatures above the single molecule
helix-globule transition temperature. For these separations, the paired
helices have strongly correlated chiralities, and for $\sigma/l=1.3$
and $1.6$, are aligned end-to-end. At separations closer than the
typical helix length, however, the molecules lose helicity and adopt
disordered configurations. At separations immediately below the typical
helix length, the disordered globular configurations of the polymers
are uncorrelated. As the separation becomes approximately equal to
the radius of gyration of the complex, however, energy anticorrelations
begin to develop between the two molecules; below this separation,
the radius of gyration of the complex remains approximately constant.
The potential of mean force is qualitatively insensitive to $\sigma/l$,
consisting of a monotonically decreasing {}``ramp'' from the initial
onset of interactions at large separations, down to the separation
at which energy correlations begin to occur between globules. Below
this separation, the PMF remains approximately constant in a broad,
flat minimum to zero separation. For the temperatures at which helical
configurations are observed, the PMF's are all strongly attractive,
as evidenced by very large negative values for the second virial coefficient
$B_{2}$, shown in Fig.~\ref{fig:B2fig}.%

We suggest that the stabilization of helical structure at moderate
separation occurs because attractive interactions between the two
molecules exert a stretching force. Stretching forces have been shown
to stabilize helix formation in other simple helix-forming models.\cite{stretching}
Correlation in chirality and helical axis can be seen as a consequence
of energy minimization. By sampling face-to-face configurations, contacts
between helices are possible at intermediate separation; by having
the same chiralities, the match-up between the faces, and hence the
number of contacts, is maximized. For $\sigma/l=1.9$, there is no
significant axial alignment, but there is a strong correlation in
chirality between helices. This suggests that face-to-face interactions
are strong enough to promote identical chirality between molecules
even when such interactions are not strong enough to hold the two
polymers in axial alignment.

\section{\label{sec:Conclusions}Conclusions}

The observed behaviors are similar to those seen in simulations of
helical homopolypeptides \cite{robinpaper} up to the separation at
which helical structure is lost. Upon close approach, simulated homopolypeptides
are observed to switch to parallel paired helical configurations,
whereas for the simple model presented here, the system prefers to
collapse into an unstructured globule. The helical structures which
the system exhibits are simply not stable enough to overcome the free
energy to be gained by formation of the collapsed structure. 

One may therefore ask how the helix can be stabilized against the
collapsed, disordered globule. We consider that the $\Theta$-point
of a polymer (the temperature at which $B_{2}=0$) is related to the
gas-liquid transition in a bulk system. The $\Theta$-point is also
the temperature below which a single polymer will begin to collapse
from an extended, disordered coil state (analogous to a gas) to a
collapsed, disordered globule state (analogous to a liquid). If we
hypothesize that the helical configuration is analogous to a bulk
crystalline phase for the polymer model (since it is an ordered state
which must be reached through a first-order transition), then we can
consider methods by which the crystalline phase may be stabilized
against the liquid phase. It has been shown \cite{Gast83,Illet95}
that, for bulk systems of particles interacting via a simple, isotropic
potential, decreasing the range of the attractive part of the potential
can decrease the stability of the liquid phase to the point of leaving
the liquid-gas critical point metastable with respect to the crystal
phase. In analogy, we suggest that by decreasing $\lambda$ in Eq.~(\ref{eq:sqwell}),
we may depress the $\Theta$-point of the system without depressing
the helix transition temperature, so that the stability of the dense
globule may decrease to the point that the observed collapse may not
occur, giving results closer to those from simulations of homopolypeptides.

This suggestion may have a basis in the chemistry of real homopolypeptides.
The strongest interaction present for these molecules is likely to
be hydrogen bonding between segments of the peptide backbone. This
hydrogen bonding is very strongly anisotropic, occurring for only
a small range of orientational alignments between peptide segments.
It has been shown that for bulk systems interacting via simple anisotropic
potentials, decreasing the range of orientations over which interactions
can occur depresses the temperature of the liquid-gas critical point.\cite{Kern2003} 

We also observe that moderate proximity between pairs of square well
homopolymers can promote the stability of helical structures. We have
suggested that this may be linked to the stabilization of helical
structure upon stretching of a single molecule, observed for other
simple helix-forming models.\cite{stretching} For pairs of molecules,
this stabilization is not enough to prevent collapse into an unstructured
globule. Experimentally, helix forming proteins aggregate to form
ordered structures with molecules in aligned, extended configurations.
Since such configurations are not seen for this model nor for simulations
using pairs of more detailed model peptides,\cite{robinpaper} it
seems that they must be stabilized by three or more body interactions
between molecules. This suggests that bulk simulations using the model
presented here may be useful.

In conclusion, we have shown that interactions between pairs of square
well homopolymers can promote helix stability, while at close proximity,
interactions cause collapse from paired helices to a single disordered
globule. We have further suggested a modification to the model (a
reduction in the range of the attractive part of the potential) which
may stabilize the helix phase against such collapse, in order to provide
better qualitative agreement with simulations of pairs of helix forming
homopolypeptides. Finally, we have suggested that the reason for the
formation of amyloid aggregate structures may lie in many-body interactions
between molecules. These results underline the need for a better understanding
of the underlying nature of the helix-disordered globule transition,
in order to gain an understanding of how interactions between molecules
can increase or decrease the relative stabilities of the various structures.

\begin{acknowledgments}
This work is supported by the EPSRC (grant reference EP/D002753/1).
\end{acknowledgments}
\bibliographystyle{apsrev}
\bibliography{swpmfs}

\end{document}